\documentclass[10pt,twocolumn,letterpaper]{article}

\usepackage{cvpr}              % To produce the CAMERA-READY version
\usepackage{algorithm}
\usepackage{algpseudocode}

\usepackage[dvipsnames]{xcolor}

\definecolor{cvprblue}{rgb}{0.21,0.49,0.74}
\usepackage[pagebackref,breaklinks,colorlinks,citecolor=cvprblue]{hyperref}

\def\paperID{*****} % *** Enter the Paper ID here
\def\confName{CVPR}
\def\confYear{2024}

\title{IMIL: Interactive Medical Image Learning Framework}

\author{
    Adrit Rao$^{1, 2}$\quad Andrea Fisher$^{1}$\quad Ken Chang$^{1}$\quad John Christopher Panagides$^{1}$\quad Katherine McNamara$^{1}$\\
    {\footnotesize \tt \{adritrao, atfisher, changk1, jpanagid, kpogrebn\}@stanford.edu} \vspace{0.01em}\\
    \and
    Joon-Young Lee$^{3}$\quad Oliver Aalami$^{1}$\\
    {\footnotesize \tt jolee@adobe.com \quad aalami@stanford.edu} \\
    $^{1}$Stanford University \quad
    $^{2}$Palo Alto High School \quad
    $^{3}$Adobe Research
}

\begin{document}
\maketitle
\begin{abstract}

Data augmentations are widely used in training medical image deep learning models to increase the diversity and size of sparse datasets. However, commonly used augmentation techniques can result in loss of clinically relevant information from medical images, leading to incorrect predictions at inference time. We propose the Interactive Medical Image Learning (IMIL) framework, a novel approach for improving the training of medical image analysis algorithms that enables clinician-guided intermediate training data augmentations on misprediction outliers, focusing the algorithm on relevant visual information. To prevent the model from using irrelevant features during training, IMIL will 'blackout' clinician-designated irrelevant regions and replace the original images with the augmented samples. This ensures that for originally mispredicted samples, the algorithm subsequently attends only to relevant regions and correctly correlates them with the respective diagnosis. We validate the efficacy of IMIL using radiology residents and compare its performance to state-of-the-art data augmentations. A 4.2\% improvement in accuracy over ResNet-50 was observed when using IMIL on only 4\% of the training set. Our study demonstrates the utility of clinician-guided interactive training to achieve meaningful data augmentations for medical image analysis algorithms .

\end{abstract}    
\section{Introduction}
\label{sec:intro}

\begin{figure}[h]
    \centering
    \includegraphics[scale=0.45]{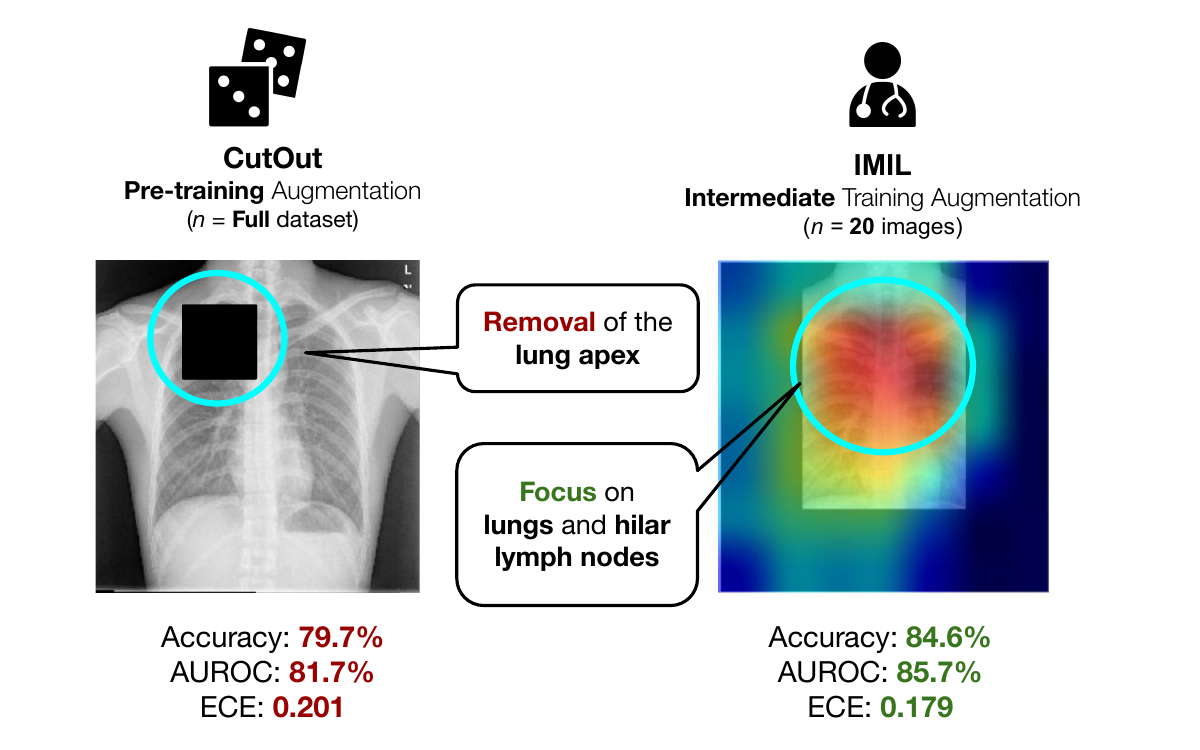}
    \caption{\textbf{IMIL versus CutOut \cite{devries2017improved}.} In this study, we propose an interactive medical image learning framework (IMIL), a callback framework that solicits clinician input to generate human-guided, intermediate training data augmentations. Compared to \textit{random} pre-training augmentations, IMIL prevents the removal of clinically relevant visual features. When IMIL is used on just \textbf{20} CXR images, it surpasses the performance and calibration of CutOut for tuberculosis classification (see Results \ref{sec:results} - IMIL + Res1).}
    \label{fig:first}
\end{figure}

The applications of computer vision to medical image analysis have been numerous in recent years \cite{esteva2021deep, chan2020deep}. This can be attributed to major advancements in deep learning and increased availability of large, open-access medical imaging datasets \cite{horien}. These algorithms have the potential to significantly improve the efficiency and accuracy of disease diagnosis in various medical imaging modalities \cite{rajpurkar2017chexnet, park2019deep, huang2020penet, rajpurkar2020appendixnet}.

As these models are translated into the clinical setting, it is important to consider the interaction between clinicians and algorithms \cite{chen2022explainable}. In addition to training algorithms for accuracy, practitioners must also ensure that a model identifies and concentrates on clinically relevant features (correct \textit{focus}) and performs consistently and dependably in various settings (\textit{reliability}). A focused algorithm should not only be able to make correct predictions but should make these predictions based on relevant regions of the medical image and at the same time, avoid spurious correlations from irrelevant parts of the image \cite{zech2018variable, winkler2019association}. Algorithms can have difficulty focusing on clinically relevant regions, as they often lack the ability to independently sort information for clinical relevance like a clinician would \cite{rao2021studying, degrave2021ai}. Furthermore, model reliability is achieved by providing confidence levels for predictions, which enable clinicians to understand the true certainty of a prediction. Standard convolutional neural networks (CNN) are often prone to overconfidence and miscalibration, creating the need for confidence calibration \cite{guo2017calibration}. Although visual explainability and calibration are important, they are not reflected in accuracy or performance-based metrics and can often go undetected. 

Image augmentations are used in medical image analysis to counteract a lack of diversity or scarcity of clinical data. Typically, these are performed before training in a randomized manner through image manipulations (e.g. crop, zoom, flip) \cite{chlap2021review}. Recent, modern augmentations include CutMix, MixUp, and CutOut \cite{yun2019cutmix, devries2017improved, zhang2017mixup}. These methods aim to improve the performance, robustness, and calibration of CNNs. Although their efficacy has been established for medical image analysis \cite{rao2023studying}, their underlying effects pose significant challenges and may necessitate further optimization. When training algorithms on clinical data, it is most efficient to retain all relevant information. However, data augmentations developed for non-medical use may run the unintended risk of removing clinically relevant information from images. Doing so may not always be directly reflected in accuracy but can significantly impact focus and visual explainability. For example, CutOut \cite{devries2017improved} (which performs randomized pre-training dropouts in the visual space), may remove the lung apex in a chest x-ray which is critical to identify tuberculosis (as shown in Figure \ref{fig:first}). The algorithm instead focuses on external and irrelevant visual features. With limited medical image datasets, it is important for the algorithm to correctly associate class labels with relevant visual indicators. Even small shifts in this relation can cause downstream effects on attention and domain shifting.

 To improve model focus and reliability, we explore the usage of clinician feedback during training, rather than before, to perform clinically meaningful augmentations on the most challenging cases for the algorithm. We aim to focus the algorithm on relevant visual information, increase the safety of augmentations, and minimize clinician annotation burden. The Interactive Medical Image Learning (IMIL) framework is a flexible callback that enables clinician-guided intermediate training data augmentations. During a set frequency in training, IMIL selects a predefined number of outliers from the training dataset based on the algorithm's worst mispredictions. These images are provided to a clinician along with the associated class activation map (CAM) \cite{selvaraju2017grad}, prediction with confidence, and ground-truth label. The clinician then re-directs the \textit{attention} of the algorithm after understanding why it is making a misprediction. To provide feedback, akin to the Google reCAPTCHA method \cite{GoogleRecaptcha2024} where users identify relevant segments within a grid on an image, the clinician selects the region of the image that the model \textit{should be} focusing on, using a similar grid overlay approach. Using this input, IMIL will perform a 'blackout' augmentation and remove all of the unselected grid regions. The newly augmented image only contains clinically relevant information that should be associated with the diagnostic label. The algorithm then re-learns the correct visual features on the most challenging outlier cases (as shown in Figure \ref{fig:first}). After feedback, the original training samples are replaced with the IMIL augmented images for subsequent training. Our study demonstrates that intermediate training augmentations based on clinician-guidance can significantly improve performance and calibration and can maximize the potential of smaller medical image datasets.

To validate the efficacy of IMIL, we perform a clinical user-study with three clinical radiology residents. The residents
provide IMIL feedback on a tuberculosis (TB) chest x-ray (CXR) dataset, which is used to create three separate IMIL-augmented algorithms. We perform a comparative analysis of these algorithms to modern augmentations (MixUp \cite{zhang2017mixup}, CutMix \cite{yun2019cutmix}, and CutOut \cite{devries2017improved}) using performance and calibration-based metrics. 

Our main \textbf{contributions} are summarized below:
\begin{enumerate}
    \item We propose a novel deep learning callback framework, \textbf{IMIL}, which incorporates clinician feedback to perform intermediate training augmentations. 
    \item We conduct a user-study on clinical residents to understand the potential of \textbf{IMIL} deployment
    \item We validate \textbf{IMIL} using deep learning algorithms trained during the user-study and compare performance against state-of-the-art modern augmentations: CutMix, MixUp, and CutOut. 
\end{enumerate}

\section{Related Work}
\label{sec:related}

\noindent\paragraph{Medical Confidence Calibration:} Various studies have shown that standard CNNs are highly prone to overconfidence which can lead to unreliable confidence estimates. Gou \textit{et al.} \cite{guo2017calibration} studied the calibration of widely used architectures, such as ResNet \cite{he2016deep}. The study found that although increased network depth and width tends to improve accuracy, it can have negative effects on calibration. Furthermore, the use of batch normalization can also reduce calibration. Calibration level is not reflected in standard performance-based metrics which can lead to this problem going undetected. However, in the general computer vision domain, various techniques have been developed to combat miscalibration. Modern augmentations such as MixUp, CutMix, and CutOut can have significant effects on calibration. In the original studies, these augmentations showed major benefits for regularization of CNNs \cite{yun2019cutmix, zhang2017mixup, devries2017improved}. Various studies have also separately validated the calibration benefits of these algorithms showing that they can yield significant improvements. They have also been applied in the medical domain for image classification and segmentation \cite{galdran2021balanced, eaton2018improving}. Additionally, a study looked at benchmarking these augmentations for various medical image modalities and revealed that they can significantly improve the performance and calibration of CNNs \cite{rao2023studying}.

\noindent\paragraph{Visual Explainability:}
Studies have shown that standard CNNs do not possess the ability to accurately rank the clinical relevance of visual features \cite{rao2021studying, saporta2022benchmarking}. In one study \cite{degrave2021ai}, the visual explainablity of a deep learning algorithm for COVID-19 detection in chest x-rays was assessed. The study found that even though performance-based metrics indicated that the algorithm was \textit{accurate}, saliency maps revealed that predictions were being made on textual indicators rather than actual pathology. This is just one example of the potential risk associated with models that lack \textit{focus}. 

\noindent\paragraph{Human-in-the-loop Training:}
 We have not identified a prior augmentation technique that incorporates clinician feedback during training. However, interactive training (human-in-the-loop) has been notably studied through active learning (AL) in the medical image domain \cite{budd2021survey}. AL often involves querying a clinical expert to label data points that will have the greatest effect on performance. This is a technique that is often leveraged when the cost of obtaining labeled data is high (time and expertise) \cite{gaillochet2023active,smailagic2018medal,yang2017suggestive}.

\section{Methods}
\label{sec:methods}

\begin{figure*}[ht]
  \centering
  \includegraphics[width=\textwidth]{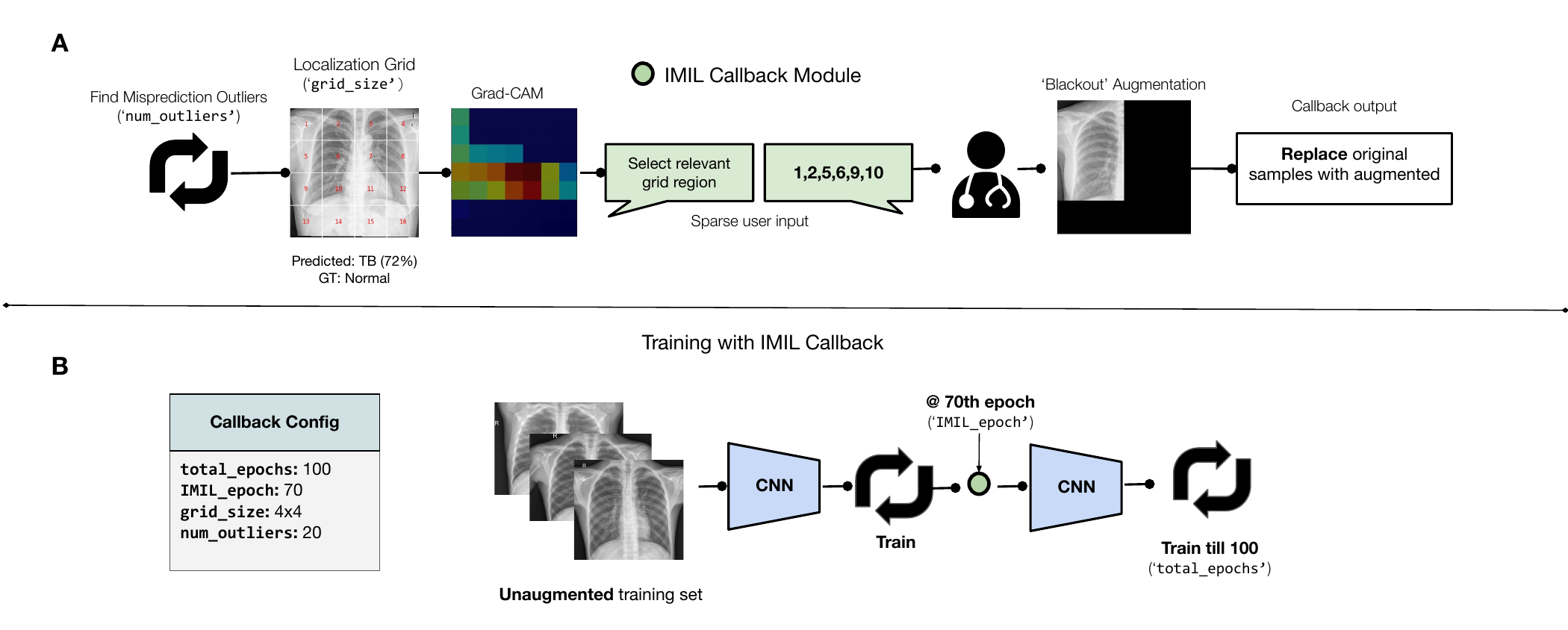}
  \caption{\textbf{Interactive Medical Image Learning (IMIL)} callback framework architecture (A) and implementation within a standard CNN training pipeline (along with callback configuration) (B). The IMIL framework allows the clinician to perform guided-augmentations ('blackout') \textit{during} training to re-focus the algorithm on clinically relevant regions. The callback consists of various configuration parameters allowing for customized usage of IMIL during training.}
  \label{fig:arch}
\end{figure*}

We perform a preliminary validation of the IMIL callback framework for TB classification in CXR images \cite{jaeger2014two}. We validate IMIL by performing a clinical user-study with residents to train three IMIL-augmented CNNs. We then compare the performance and confidence calibration of IMIL to state-of-the-art modern augmentations: CutMix \cite{yun2019cutmix}, MixUp \cite{zhang2017mixup}, and CutOut \cite{devries2017improved}. As follows is a description of the callback architecture (\ref{sec:arch}), modern data augmentations (\ref{sec:mod}), and study design (\ref{sec:design}). 

\subsection{IMIL Architecture}
\label{sec:arch}

The overall architecture of the IMIL callback module is shown in Figure \ref{fig:arch}A and the usage of IMIL within a training pipeline for a CNN (as done in our study) is shown in Figure \ref{fig:arch}B. Figure \ref{fig:arch}B also shows the IMIL callback configuration used within our study. Each of the parameters are customizable within the framework and consist of \texttt{num\_outliers}, \texttt{IMIL\_epoch}, and \texttt{grid\_size}. As follows are details regarding each step of the feedback loop.

\noindent\paragraph{Misprediction Outlier Selection} Based on the predefined \texttt{num\_outliers} parameter in the callback configuration, IMIL finds the 'most significant' mispredictions on which clinician feedback will be obtained. At the \texttt{IMIL\_epoch} at which the callback is called, IMIL makes predictions across the full training set. IMIL then compiles a list of mispredictions by comparing ground-truth labels to model predictions. The mispredictions are then sorted in descending order based on confidence. The final outliers (\textit{n} = \texttt{num\_outliers}) are then chosen from the \textit{highest confidence mispredictions}. This process helps direct human feedback towards the most challenging cases for the model. As such, directing human feedback to shift attention on hallucinations can potentially have more significant effects on downstream feature extraction compared to a random outlier selection strategy.

\noindent\paragraph{User Interpretation}
For each outlier, the clinician is provided with a visual dialogue during training to interpret the misprediction and then provide feedback. Before the clinician is prompted for feedback, they are provided with information to inform their decision-making: the original image, CAM heat-map \cite{selvaraju2017grad}, predicted label (with confidence), and ground-truth (as shown after misprediction selection in Figure \ref{fig:arch}A). The clinician should interpret all datapoints to understand the cause for the misprediction. For example, the heat-map can direct the clinicians attention to the region of the image where the model extracted features. The predicted label and associated confidence can help the clinician determine how these features are being mapped in the CNN. As the clinician becomes more familiar with this interpretation flow, they may start to notice biases in the model's focus (e.g. consistent attention towards the soft tissues instead of lung in CXR) and thus tailor their feedback to better correct the model.

\noindent\paragraph{Localization Grid Feedback}
Clinician feedback is obtained using a localization grid overlaid on the original image. The grid is made up of equally sized squares of modifiable dimensions (based on \texttt{grid\_size}). The clinician selects relevant squares on the grid overlay that correspond to clinically relevant visual features (as shown in the sparse user input stage in Figure \ref{fig:arch}A). This step re-focuses the model and prevents it from falsely associating the ground-truth class label to irrelevant features. In the current prototype implementation of IMIL, this is done by assigning each grid region a number and mapping to a console input. However, the underlying callback function can be applied to a variety of different interactive feedback interfaces.

\noindent\paragraph{Callback Output Augmentation}
The region-based sparse input is used to generate the augmented output. IMIL removes unselected regions ('blackout') from the image (similar to \textit{visual dropout} \cite{devries2017improved}), and retains the selected regions. The original training sample is then replaced with the augmented IMIL output at the specified \texttt{IMIL\_epoch}. The replacement of the original image during training allows the algorithm to re-associate the ground-truth label with relevant visual features for the remainder of training epochs. After replacment, training continues (from the 70th to 100th epoch; Figure \ref{fig:arch}B).

\subsection{Modern Data Augmentations}
\label{sec:mod}
We compare IMIL against MixUp \cite{zhang2017mixup}, CutMix \cite{yun2019cutmix}, and CutOut \cite{devries2017improved} which have been widely studied and applied for medical image analysis. Out of these augmentations, IMIL's 'blackout'-like augmentation is similar to CutOut. However, instead of random visual dropout it is clinician-guided and performed during training on a limited sample size. As follows are brief details surrounding the formulation for each augmentation. 

\noindent\paragraph{MixUp}

Zhang \textit{et al.} \cite{zhang2017mixup} proposed MixUp as a modern augmentation technique for training neural networks on a \textit{blend} between a pair of images and labels based on convex combinations. MixUp has demonstrated benefits in terms of increasing robustness of neural networks when learning from corrupt labels and adversarial examples. The original formulation of MixUp from the original paper \cite{zhang2017mixup} is:

\begin{equation}
\begin{array}{l}
\tilde{x}=\lambda x_{i}+(1-\lambda) x_{j} \\
\tilde{y}=\lambda y_{i}+(1-\lambda) y_{j},
\end{array}
\end{equation}
where $\boldsymbol{x}_{i,}, \boldsymbol{y}_i$ are raw randomly sampled input vectors and $\boldsymbol{x}_{j}, \boldsymbol{y}_j$ are the corresponding one-hot label encodings. $\lambda$ are values in the range [0, 1] which are randomly sampled from the Beta distribution for each augmented example.

% \jl{define and explain $\lambda$.}

\noindent\paragraph{CutMix}

Yun \textit{et al.} \cite{yun2019cutmix} introduced CutMix, an augmentation built upon the original formulation of MixUp and the idea of combining samples. CutMix removes a patch from an image and swaps it for a region of another image generating a locally natural unseen sample. The formulation for CutMix is as follows:

\begin{equation}
\begin{array}{c}
\tilde{x}=M x_{i}+(1-M) x_{j} \\
\tilde{y}=\mu y_{i}+(1-\mu) y_{j},
\end{array}
\end{equation}
where $M$ indicates the binary mask used to perform the cutout and fill-in operation from two randomly drawn images. $\mu$ are values (in [0,1]) randomly drawn from the Beta distribution. 

% and $\mu$ 
% \jl{define and explain $M$ and $\mu$. I think $\lambda$ above and $\mu$ here have different definitions.}

\noindent\paragraph{CutOut} 

This technique was proposed by DeVries \textit{et al.} \cite{devries2017improved} and is a simple augmentation technique for improving the regularization of CNNs. CutOut was formulated based on the idea of extending dropout \cite{hinton2012improving} to a spatial prior in the input space. CutOut performs occlusions of an input image similar to the idea proposed in \cite{bengio2011deep}. Rather than partially occluding portions of an image \cite{bengio2011deep}, CutOut performs fixed-size zero-masking to fully obstruct a random location of an image. CutOut differentiates from dropout as it is an augmentation technique and visual features are dropped at the input stage of the CNN whereas in dropout, this occurs in intermediate layers. The goal of CutOut is to not only improve regularization of CNNs but improve robustness to occluded samples in real-world applications.

\subsection{Study Design}
\label{sec:design}

As follows is a description of each component of the user-study for feedback-guided augmentations. The dataset used for this study is described in \ref{sec:dataset}, the configuration of IMIL in \ref{sec:config}, the CNN architectures used in \ref{sec:archs}, training details in \ref{sec:details}, clinical user-study structure in \ref{sec:user}, and lastly the evaluation metrics used to measure performance and calibration of the models in \ref{sec:perfcal}.

\subsubsection{Tuberculosis Dataset}
\label{sec:dataset}

To validate the efficacy of IMIL, we focused on TB diagnosis in CXR. TB is a highly prevalent lung condition resulting in more than 1 million deaths per year worldwide; thus, significant international attention has been paid to prompt diagnosis and treatment of the disease \cite{glaziou}. Chest x-ray is an effective and cost-efficient modality for pulmonary TB diagnosis, making it a vital clinical tool in low-resource settings where TB is most prevalent \cite{vancleeff}. Moreover, labeled CXR datasets are available, and AI for TB diagnosis in CXR has been validated in previous research \cite{geric}. The dataset selected for this study was from the U.S. National Library of Medicine (NIH) Shenzhen No. 3 People’s Hospital in China \cite{jaeger2014two}. The dataset consists of 662 frontal chest x-ray (CXR) images labeled with each respective patient’s TB positive or negative diagnosis. The limited sample size helps demonstrate the effective use of data through IMIL's feedback-based augmentations. The dataset has a class distribution of 49\% positive and 51\% TB negative. In terms of demographics, the dataset only included sex: 69\% male and 31\% female. The models in our study were trained on 80\% (\textit{n} = 530) of the dataset and evaluated on 20\% (\textit{n} = 132).   

\subsubsection{IMIL Configuration}
\label{sec:config}

For our study, we configured the IMIL callback to launch at the 70th epoch (\texttt{IMIL\_epoch}) of training. We also used an IMIL grid size of 4x4 (\texttt{grid\_size}). Lastly, we use 20 for the \texttt{num\_outliers} parameter. In summary, IMIL will function by sequentially providing the clinician with 20 misprediction outliers (which accounts for 3.77\% of the training dataset) at the 70th epoch and after feedback is provided, training will continue till the 100th epoch using the dataset with the newly augmented samples.

\subsubsection{Model Architecture}
\label{sec:archs}

The CNN architecture used in this study was ResNet-50 \cite{he2016deep} which is widely used for medical image analysis, notably through transfer learning \cite{nguyen2019deep,hossain2022transfer,showkat2022efficacy}. The ResNet model was implemented based on the standard Tensorflow Keras \cite{tensorflow2015-whitepaper} applications plugin \footnote{\url{https://keras.io/api/applications}}. The implementations of the MixUp, CutMix, and CutOut followed the original formulations \footnote{\href{https://github.com/ayulockin/DataAugmentationTF}{DataAugmentationTF Repository}}. For CutOut we used a 50x50 pixel mask size.

\subsubsection{Training Details}
\label{sec:details}

Each of the seven ResNet-50 models (baseline, CutMix, CutOut, MixUp, IMIL + Resident 1/2/3) were trained for 100 epochs. Each input is trained to two output logits corresponding to the normal and TB class labels. Experiments are performed using the stochastic gradient descent (SGD) optimizer \cite{kingma2014adam}, a batch size of 64, and learning rate 0.001. All input images are uniformly scaled to 224x224.

\subsubsection{Clinical User-Study}
\label{sec:user}

Our clinical user-study to validate the efficacy of IMIL was performed with three radiology residents at Stanford Medicine. These trainees will be referred to as residents 1, 2, and 3. Resident 1 is a PGY-2 diagnostic radiology resident, resident 2 is a PGY-2 interventional radiology resident, and resident 3 is a PGY-1 interventional radiology resident. Three separate ResNet-50 models were trained for each resident. The total time of interaction between each resident and the feedback mechanism lasted less than 30 minutes. Before each training session, two main components of IMIL were described to the resident. The first component is how to interpret the predicted label and CAM output jointly to understand where the model was focusing on to reach its prediction. The second component described is the how to perform the region selection and how the grid overlay functions. The resident was informed about the objective of the feedback: to shift the model's focus from irrelevant to clinically significant features that correlate to the ground-truth label. The residents were instructed to prioritize a single specific region for feedback, despite the possibility of multiple relevant areas.

\subsubsection{Evaluation Metrics}
\label{sec:perfcal}

Below are descriptions of the statistical metrics used to validate the CNN. The first two metrics are used to validate the performance of the algorithm. The third metric is used to evaluate confidence calibration which is not reflected in performance-based metrics. 

\noindent\paragraph{Accuracy and AUROC}

To evaluate the performance of the algorithm, we use the test set accuracy and area under the receiver operating characteristic (AUROC). The accuracy measures the fraction of predictions that were made correctly across the test set after training. AUROC is a robust measure of the ability for the binary classifier to discriminate between class labels \cite{hanley1982meaning}.

\noindent\paragraph{Expected Calibration Error (ECE)}

The ECE is commonly used to quantify the level of confidence calibration for algorithms. This approach provides a scalar summary statistic of calibration by grouping a model's predictions into equally spaced bins (\textit{B}). The weighted average of the difference between accuracy and confidence across the bins is outputted. The formulation of ECE from \cite{guo2017calibration} is shown:

\begin{equation}
\mathrm{ECE}=\sum_{b=1}^{B} \frac{n_{b}}{N}|\operatorname{acc}(b)-\operatorname{conf}(b)|,
\end{equation}

where \textit{n} represents the number of samples. Gaps in calibration or miscalibration are represented by the difference between $\operatorname{acc}$ and $\operatorname{conf}$. In our study, we use a bin size of 15.

\section{Results}
\label{sec:results}

\begin{figure*}[ht!]
    \centering
    \begin{minipage}{0.48\textwidth}
        \centering
        \includegraphics[width=\linewidth]{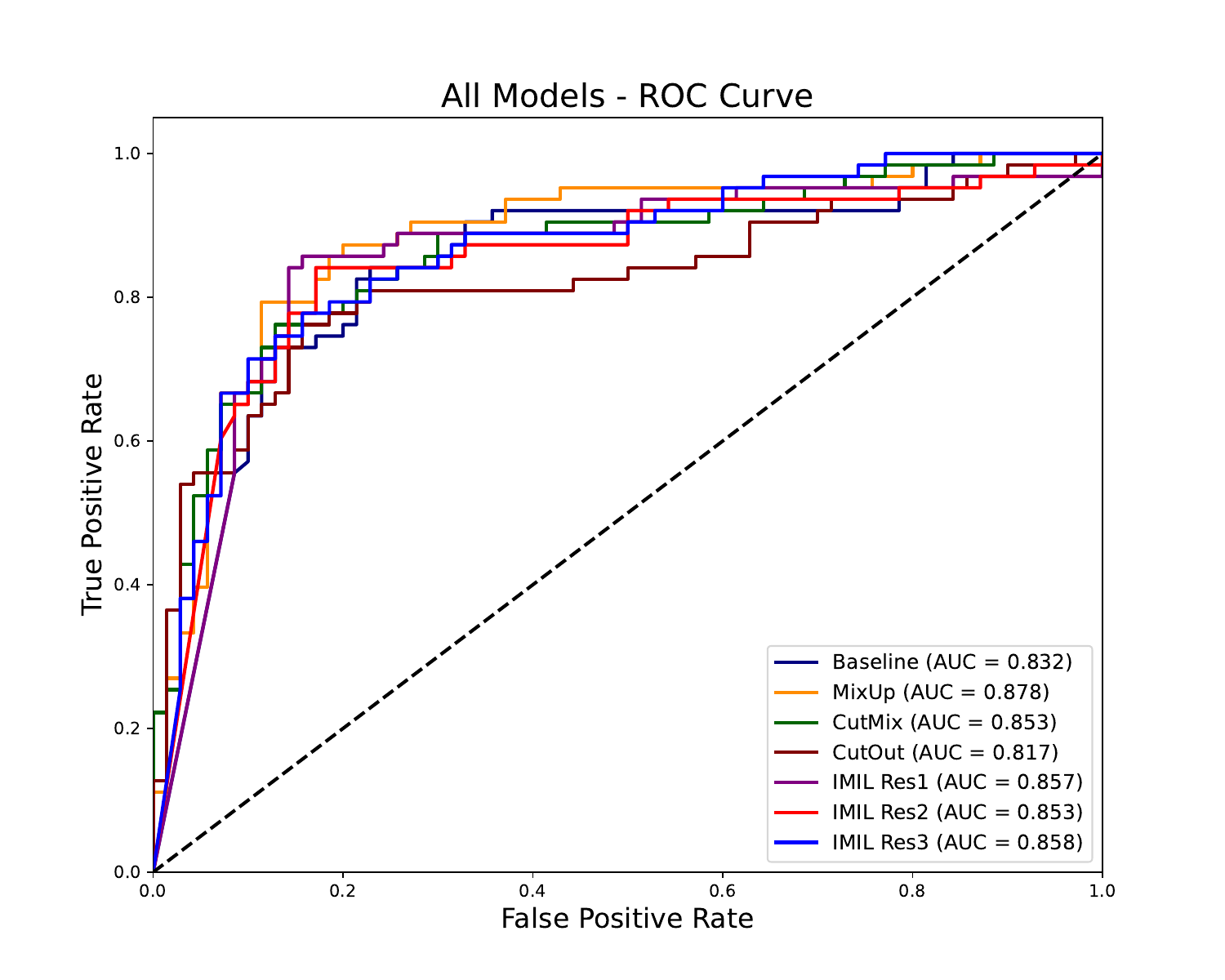}
        \caption{\textbf{ROC Curve.} The curve represents the comparison of AUC between the baseline, modern augmentations (MixUp \cite{zhang2017mixup}, CutMix \cite{yun2019cutmix}, and CutOut \cite{devries2017improved}), as well as the three IMIL-trained resident algorithms (IMIL Res1, Res2, and Res3).}
        \label{fig:roc}
    \end{minipage}\hfill
    \begin{minipage}{0.49\textwidth}
        \centering
        \includegraphics[width=\linewidth]{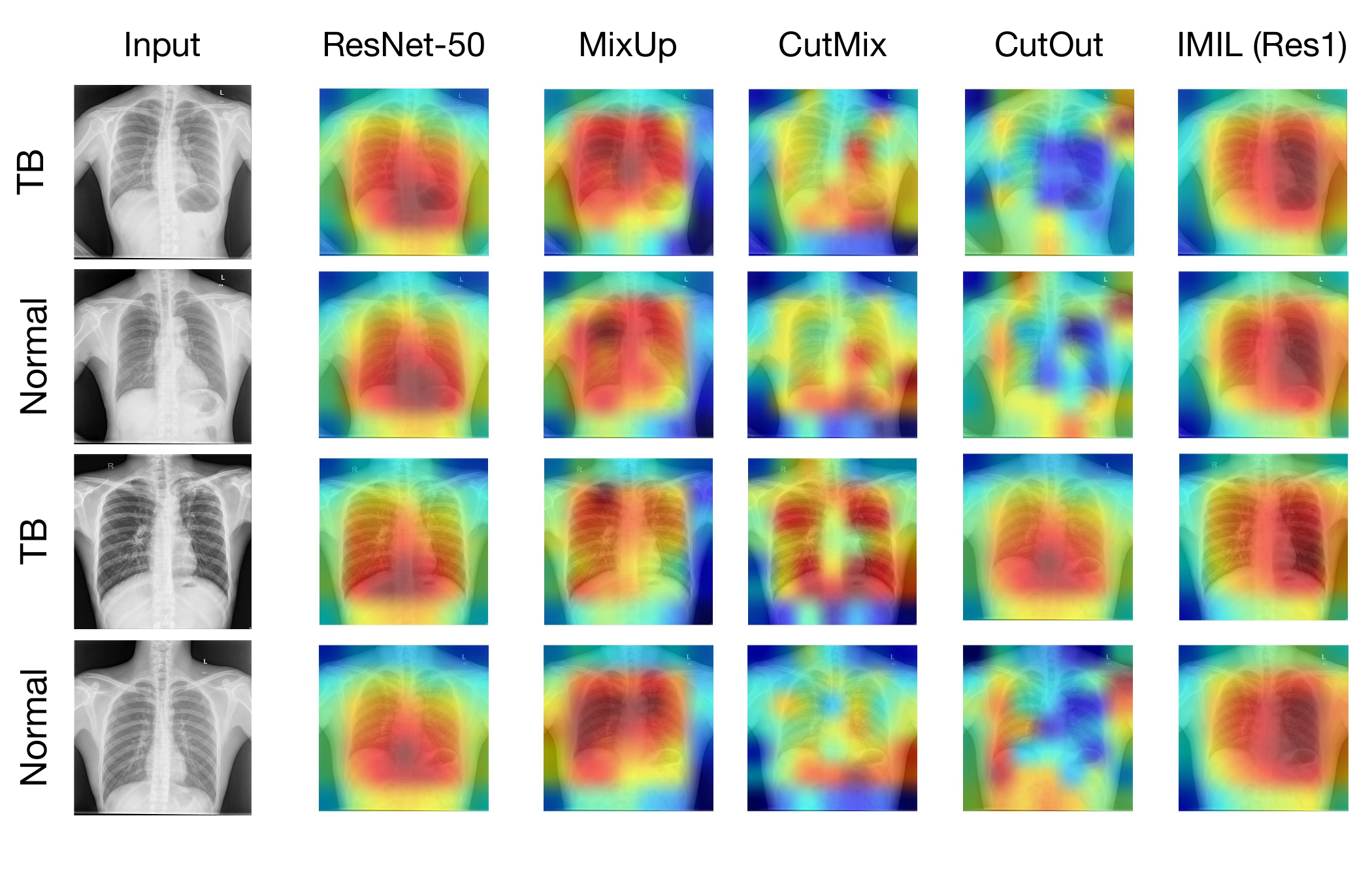}
        \caption{\textbf{CAM Visualizations.} Shown above are CAM heat-maps comparing the baseline with the modern augmentations (MixUp \cite{zhang2017mixup}, CutMix \cite{yun2019cutmix}, and CutOut \cite{devries2017improved}) as well as IMIL (for Resident 1 which had the highest test set accuracy).}
        \label{fig:visualizations}
    \end{minipage}
\end{figure*}

\begin{table*}
\centering
\begin{tabular}{|c|c|c|c|}
\hline
\textbf{Model}    & \textbf{Accuracy} & \textbf{AUROC} & \textbf{Expected Calibration Error (ECE)} \\ \hline
Baseline (ResNet-50 \cite{devries2017improved})         & 0.804             & 0.832          & 0.2                                       \\ \hline
CutMix \cite{yun2019cutmix}            & 0.812 (+0.008)    & 0.853 (+0.021) & 0.15 (-0.05)                              \\ \hline
CutOut \cite{devries2017improved}           & 0.797 (-0.007)    & 0.817 (-0.015) & 0.201 (+0.001)                            \\ \hline
MixUp \cite{zhang2017mixup}             & 0.834 (+0.03)     & 0.878 (+0.046) & 0.161 (-0.039)                            \\ \hline
IMIL + Resident 1 & \textbf{0.846 (+0.042)}    & 0.857 (+0.025) & 0.179 (-0.021)                            \\ \hline
IMIL + Resident 2 & 0.835 (+0.031)    & 0.853 (+0.021) & \textbf{0.175 (-0.025)}                            \\ \hline
IMIL + Resident 3 & 0.824 (+0.02)     & \textbf{0.858 (+0.026)} & 0.183 (-0.017)                            \\ \hline
\end{tabular}
\caption{\textbf{Validation statistics of baseline, augmentations, and IMIL framework (for the three residents).} Statistics consist of test set accuracy, AUROC score, and Expected Calibration Error (ECE) (\textit{M = 15 bins}) for the TB CXR dataset. Bold values represent the best performing IMIL resident model.}
\label{table:calibration_and_performance}
\end{table*}

\begin{figure*}[h]
  \centering
  \includegraphics[width=\textwidth]{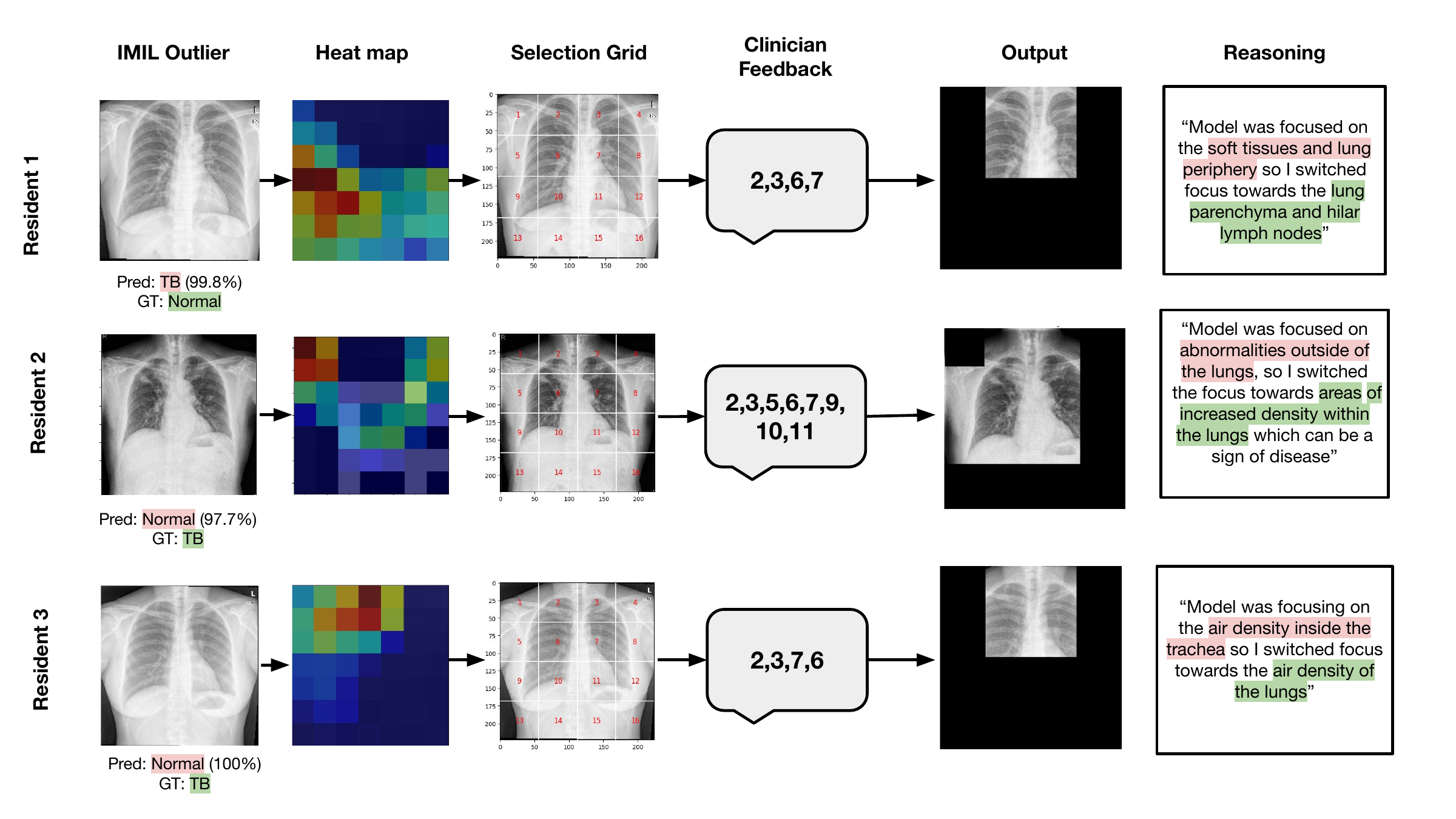}
  \caption{\textbf{IMIL Feedback Samples}. End-to-end interaction samples are shown above with the IMIL outlier and the predicted label (with confidence) and the ground-truth. Next to that is the heat-map and selection grid which the resident uses to provide feedback in the form of a numerical input. This is then used to perform the 'blackout' augmentation. Each resident also provided clinical reasoning for their feedback decisions in the samples shown.}
  \label{fig:interaction}
\end{figure*}

IMIL data augmentations yielded improvements in performance and calibration over baseline and CutOut. Performance-based metrics are discussed in \ref{sec:performance} and calibration in \ref{sec:calibration}. In \ref{sec:visualizations}, CAM visualizations from the different algorithms are presented as well as IMIL interaction samples from each clinical resident to assess usability.

\subsection{Performance}
\label{sec:performance}

% \begin{figure}[h!]
%     \centering
%     \includegraphics[width=\linewidth]{figs/ROC.pdf}
%     \caption{\textbf{ROC Curve.} The curve represents the comparison of AUC between the baseline, modern augmentations (MixUp \cite{zhang2017mixup}, CutMix \cite{yun2019cutmix}, and CutOut \cite{devries2017improved}), as well as the three IMIL-trained resident algorithms (IMIL Res1, Res2, and Res3).}
%     \label{fig:roc}
% \end{figure}

The performance-based metrics for the different ResNet-50 \cite{he2016deep} data augmentation variants are shown in Table \ref{table:calibration_and_performance} in the first two columns (accuracy and AUROC). The baseline ResNet-50 model demonstrated accuracy of 80.4\% and AUROC of 83.2\% on the CXR test set (Row 1). Of the modern data augmentations, MixUp presented the most significant performance improvements in accuracy and AUROC with accuracy of 83.4\% (+3\% over ResNet-50) and AUROC of 87.8\%. CutMix showed less significant improvements in performance. CutOut reduced accuracy to 79.7\% (-0.7\% compared to ResNet-50) and AUROC score to 81.7\% (-1.5\% compared to ResNet-50). Rows 5-7 present the performance of the IMIL-augmented models trained separately for each resident's feedback loop. All IMIL models present performance improvements over baseline with more consistency observed in AUROC compared to accuracy. The first IMIL model (with Resident 1) presented the most significant improvement in accuracy compared to baseline and all other augmentations at 84.6\% (+4.2\% over ResNet-50). The third IMIL model (with Resident 3) showed the highest improvement in AUROC at 85.8\% (+2.6\% over ResNet-50). A summary of AUROC performance is also shown in the Figure \ref{fig:roc} ROC Curve.

\subsection{Calibration}
\label{sec:calibration}

The ECE scores for each algorithm are shown in the last column of Table \ref{table:calibration_and_performance}. Lower ECE scores represent higher levels of confidence calibration. The baseline ResNet-50 had an ECE score of 0.2. Every augmented model decreased the ECE except for CutOut, which had almost no effect on it (+0.001 compared to baseline). The most significant decrease in ECE was observed in CutMix at 0.15 (-0.05 less than ResNet-50). MixUp also had significant calibration improvements with an ECE of 0.179 (-0.021 less than ResNet-50). Considering the limited outlier count, IMIL also presented significant and consistent improvements in ECE, although it did not outperform CutMix and CutOut in terms of calibration. The most significant improvement was observed in the second IMIL model (with Resident 2), which received an ECE of 0.175 (-0.025 below ResNet-50).  

\subsection{Visualizations}
\label{sec:visualizations}

% \begin{figure}[h!]
%     \centering
%     \includegraphics[width=\linewidth]{figs/CAM.pdf}
%     \caption{\textbf{CAM Visualizations.} Shown above are CAM heat-maps comparing the baseline with the modern augmentations (MixUp \cite{zhang2017mixup}, CutMix \cite{yun2019cutmix}, and CutOut \cite{devries2017improved}) as well as IMIL (for Resident 1 which had the highest test set accuracy). }
%     \label{fig:visualizations}
% \end{figure}

To gauge the effect of the various augmentations on explainability, we generate CAM visualizations on four random test set samples images (two TB and two normal) as shown in Figure \ref{fig:visualizations}. We examine visualizations for the IMIL Resident 1 model as it yielded the highest test set accuracy. As shown in the figure, there are significant variations in the model's CAM heat-maps based on the augmentation technique used. The heat-map for the baseline seems to be distributed across the lung and hilar lymph node region region, with greatest attention paid to the lower and middle lung zones. MixUp pays preferential attention to the upper zones. CutMix has a patchy heterogenous focus on the lungs. Visually, CutOut appears to focus the least reliably on lung fields, in some cases focusing preferentially on shoulder and thorax soft tissues. In contrast, Res1 IMIL heat-maps are well-focused on all of the lung fields and hilar lymph nodes. 

We also present three notable sample interactions between the resident and IMIL to demonstrate the feedback mechanism in Figure \ref{fig:interaction}. The IMIL outlier heat-map and selection grid are presented to the resident along with the predicted label, confidence, and ground-truth.  Clinician-selected grid regions are shown along with the augmented output. Each resident also provides a reasoning in the form ``The model was focused on .... so I switched the focus towards ....'' to understand their decision-making process behind the feedback and effort to shift focus. 

\section{Discussion}
\label{sec:discussion}

The main objective of this study is developing a human-in-the-loop learning framework and validating the efficacy of clinical feedback during training. We choose a CutOut-like “blackout” augmentation for two reasons. First, within IMIL, asking clinicians to select relevant regions of the image to shift the model's focus is highly intuitive; refocusing the model on lymphadenopathy or a lung opacity as opposed to clavicle or a gastric bubble was easily understood by all participants. Second, it allows for a clear validation of the framework in comparison with the CutOut baseline: random blackout vs. human-guided blackout. Although this type of augmentation was chosen with the usability of the framework in mind, it is worth considering how human feedback could be integrated into a CutMix or MixUp style of augmentation.  CutMix and MixUp  seemed to outperform CutOut in our investigation, which suggests that the benefits of IMIL clinician input adapted to these two paradigms could deliver even greater performance and calibration improvements.  Future work may investigate whether IMIL could drive accuracy gains in non-CNN models as well.

In the current study, each of the three IMIL experiments were done on a single resident respectively. Each experiment shows performance and calibration improvements over baseline and CutOut. Although the improvements for AUROC are fairly consistent, we do notice more significant variations in accuracy between residents. 

Avenues of future research should include variations on dataset/sample size and combinations of clinical participants. We validated IMIL on a single smaller scale dataset where limited feedback (20 outliers) had a significant effect. It is worth studying the effect of the IMIL outlier count on performance, as too few outliers can be insignificant, but too many can overly corrupt the data and shift the domain. It remains to be determined what the optimal ratio of clinician feedback-to-dataset size is when expanding this framework to significantly larger datasets. We hope to apply IMIL to different imaging modalities (CT, MRI) and different disease entities to validate its utility across clinical use cases. We also plan to conduct more user-studies on diagnosticians in various stages of practice (all participating residents in this study have significant familiarity in developing AI for radiology applications) to ensure our framework remains intuitive for a wide swathe of clinical users. For large datasets, it may be necessary to combine feedback from several clinicians into the same model.  We would need to test whether a multi-radiologist trained algorithm produces valid results, or whether inter-radiologist variability would serve as a confounder.

Finally, we only launch the IMIL callback at a single instance during training (70th epoch), but the callback may be employed multiple times during training. Therefore, when training models for a greater number of epochs, we could compare the timing and number of IMIL occurrences during training (e.g. comparing a single feedback loop for 20 outliers versus having two loops for 10 outliers each). Optimal grid size is also unknown and may depend on the disease and imaging modality. In some cases, having more grid regions to choose from may improve performance and allow the clinician to provide even more detailed feedback.

From a clinical perspective, IMIL holds the potential to be easily integrated into a clinical radiologist's workflow. Because AI is not yet so diagnostically reliable (and trusted by providers, patients, and the public) that it can be used in absence of a radiologist \cite{bahl}, most AI tools for triaging images to improve radiologist efficiency instead of outright replacing radiologists \cite{mellothoms}. Humans therefore still oversee final imaging reports. Discrepancies between model and radiologist interpretations could therefore be fed back into a training loop along with IMIL-style image augmentations to improve subsequent model performance. As patient populations, disease presentations, and diagnostic criteria evolve over time, ensuring AI solutions for radiologists remain flexible will be paramount \cite{pianykh}. Interactive learning for medical imaging can therefore be critical in training better out-of-the-box models for image analysis and easily adapting these models to best suit the specific needs of the hospitals and patient populations in which they are deployed.

\section{Conclusion}
\label{sec:conclusion}

Using interactive clinician-guided intermediate training feedback from three radiology residents, the IMIL callback framework and data augmentation demonstrates significant and reliable improvements in accuracy, AUROC, and calibration compared with ResNet-50 baseline and CutOut data augmentation for pulmonary tuberculosis classification on chest x-rays. The IMIL framework thus holds great potential for improving performance of computer vision models for medical imaging tasks hampered by small dataset size. Future investigation is needed to elucidate optimal outlier-to-dataset ratio, epoch timing, and image masking techniques, and to validate the model for use among different practitioners, patient populations, and imaging modalities.
{
    \small
    \bibliographystyle{ieeenat_fullname}
    \bibliography{main}

\begin{thebibliography}{43}
\providecommand{\natexlab}[1]{#1}
\providecommand{\url}[1]{\texttt{#1}}
\expandafter\ifx\csname urlstyle\endcsname\relax
  \providecommand{\doi}[1]{doi: #1}\else
  \providecommand{\doi}{doi: \begingroup \urlstyle{rm}\Url}\fi

\bibitem[Abadi et~al.(2015)Abadi, Agarwal, Barham, Brevdo, Chen, Citro, Corrado, Davis, Dean, Devin, Ghemawat, Goodfellow, Harp, Irving, Isard, Jia, Jozefowicz, Kaiser, Kudlur, Levenberg, Man\'{e}, Monga, Moore, Murray, Olah, Schuster, Shlens, Steiner, Sutskever, Talwar, Tucker, Vanhoucke, Vasudevan, Vi\'{e}gas, Vinyals, Warden, Wattenberg, Wicke, Yu, and Zheng]{tensorflow2015-whitepaper}
Mart\'{i}n Abadi, Ashish Agarwal, Paul Barham, Eugene Brevdo, Zhifeng Chen, Craig Citro, Greg~S. Corrado, Andy Davis, Jeffrey Dean, Matthieu Devin, Sanjay Ghemawat, Ian Goodfellow, Andrew Harp, Geoffrey Irving, Michael Isard, Yangqing Jia, Rafal Jozefowicz, Lukasz Kaiser, Manjunath Kudlur, Josh Levenberg, Dandelion Man\'{e}, Rajat Monga, Sherry Moore, Derek Murray, Chris Olah, Mike Schuster, Jonathon Shlens, Benoit Steiner, Ilya Sutskever, Kunal Talwar, Paul Tucker, Vincent Vanhoucke, Vijay Vasudevan, Fernanda Vi\'{e}gas, Oriol Vinyals, Pete Warden, Martin Wattenberg, Martin Wicke, Yuan Yu, and Xiaoqiang Zheng.
\newblock {TensorFlow}: Large-scale machine learning on heterogeneous systems, 2015.
\newblock Software available from tensorflow.org.

\bibitem[Bahl(2022)]{bahl}
Manisha Bahl.
\newblock Artificial intelligence in clinical practice: implementation considerations and barriers.
\newblock \emph{Journal of Breast Imaging}, 4\penalty0 (6):\penalty0 632--639, 2022.

\bibitem[Bengio et~al.(2011)Bengio, Bastien, Bergeron, Boulanger-Lewandowski, Breuel, Chherawala, Cisse, C{\^o}t{\'e}, Erhan, Eustache, et~al.]{bengio2011deep}
Yoshua Bengio, Fr{\'e}d{\'e}ric Bastien, Arnaud Bergeron, Nicolas Boulanger-Lewandowski, Thomas Breuel, Youssouf Chherawala, Moustapha Cisse, Myriam C{\^o}t{\'e}, Dumitru Erhan, Jeremy Eustache, et~al.
\newblock Deep learners benefit more from out-of-distribution examples.
\newblock In \emph{Proceedings of the Fourteenth International Conference on Artificial Intelligence and Statistics}, pages 164--172. JMLR Workshop and Conference Proceedings, 2011.

\bibitem[Budd et~al.(2021)Budd, Robinson, and Kainz]{budd2021survey}
Samuel Budd, Emma~C Robinson, and Bernhard Kainz.
\newblock A survey on active learning and human-in-the-loop deep learning for medical image analysis.
\newblock \emph{Medical image analysis}, 71:\penalty0 102062, 2021.

\bibitem[Chan et~al.(2020)Chan, Samala, Hadjiiski, and Zhou]{chan2020deep}
Heang-Ping Chan, Ravi~K Samala, Lubomir~M Hadjiiski, and Chuan Zhou.
\newblock Deep learning in medical image analysis.
\newblock \emph{Deep Learning in Medical Image Analysis}, pages 3--21, 2020.

\bibitem[Chen et~al.(2022)Chen, Gomez, Huang, and Unberath]{chen2022explainable}
Haomin Chen, Catalina Gomez, Chien-Ming Huang, and Mathias Unberath.
\newblock Explainable medical imaging ai needs human-centered design: guidelines and evidence from a systematic review.
\newblock \emph{NPJ digital medicine}, 5\penalty0 (1):\penalty0 156, 2022.

\bibitem[Chlap et~al.(2021)Chlap, Min, Vandenberg, Dowling, Holloway, and Haworth]{chlap2021review}
Phillip Chlap, Hang Min, Nym Vandenberg, Jason Dowling, Lois Holloway, and Annette Haworth.
\newblock A review of medical image data augmentation techniques for deep learning applications.
\newblock \emph{Journal of Medical Imaging and Radiation Oncology}, 65\penalty0 (5):\penalty0 545--563, 2021.

\bibitem[DeGrave et~al.(2021)DeGrave, Janizek, and Lee]{degrave2021ai}
Alex~J DeGrave, Joseph~D Janizek, and Su-In Lee.
\newblock Ai for radiographic covid-19 detection selects shortcuts over signal.
\newblock \emph{Nature Machine Intelligence}, 3\penalty0 (7):\penalty0 610--619, 2021.

\bibitem[DeVries and Taylor(2017)]{devries2017improved}
Terrance DeVries and Graham~W Taylor.
\newblock Improved regularization of convolutional neural networks with cutout.
\newblock \emph{arXiv preprint arXiv:1708.04552}, 2017.

\bibitem[Eaton-Rosen et~al.(2018)Eaton-Rosen, Bragman, Ourselin, and Cardoso]{eaton2018improving}
Zach Eaton-Rosen, Felix Bragman, Sebastien Ourselin, and M~Jorge Cardoso.
\newblock Improving data augmentation for medical image segmentation.
\newblock 2018.

\bibitem[Esteva et~al.(2021)Esteva, Chou, Yeung, Naik, Madani, Mottaghi, Liu, Topol, Dean, and Socher]{esteva2021deep}
Andre Esteva, Katherine Chou, Serena Yeung, Nikhil Naik, Ali Madani, Ali Mottaghi, Yun Liu, Eric Topol, Jeff Dean, and Richard Socher.
\newblock Deep learning-enabled medical computer vision.
\newblock \emph{NPJ digital medicine}, 4\penalty0 (1):\penalty0 5, 2021.

\bibitem[Gaillochet et~al.(2023)Gaillochet, Desrosiers, and Lombaert]{gaillochet2023active}
M{\'e}lanie Gaillochet, Christian Desrosiers, and Herv{\'e} Lombaert.
\newblock Active learning for medical image segmentation with stochastic batches.
\newblock \emph{Medical Image Analysis}, 90:\penalty0 102958, 2023.

\bibitem[Galdran et~al.(2021)Galdran, Carneiro, and Gonz{\'a}lez~Ballester]{galdran2021balanced}
Adrian Galdran, Gustavo Carneiro, and Miguel~A Gonz{\'a}lez~Ballester.
\newblock Balanced-mixup for highly imbalanced medical image classification.
\newblock In \emph{International Conference on Medical Image Computing and Computer-Assisted Intervention}, pages 323--333. Springer, 2021.

\bibitem[Geric et~al.(2023)Geric, Qin, Denkinger, Kik, Marais, Anjos, David, Ahmad~Khan, and Trajman]{geric}
C Geric, ZZ Qin, CM Denkinger, SV Kik, B Marais, A Anjos, PM David, F Ahmad~Khan, and A Trajman.
\newblock The rise of artificial intelligence reading of chest x-rays for enhanced tb diagnosis and elimination.
\newblock \emph{The International Journal of Tuberculosis and Lung Disease}, 27\penalty0 (5):\penalty0 367--372, 2023.

\bibitem[Glaziou et~al.(2013)Glaziou, Falzon, Floyd, and Raviglione]{glaziou}
Philippe Glaziou, Dennis Falzon, Katherine Floyd, and Mario Raviglione.
\newblock Global epidemiology of tuberculosis.
\newblock In \emph{Seminars in respiratory and critical care medicine}, pages 003--016. Thieme Medical Publishers, 2013.

\bibitem[{Google}(2024)]{GoogleRecaptcha2024}
{Google}.
\newblock About recaptcha.
\newblock \url{https://www.google.com/recaptcha/about/}, 2024.
\newblock Accessed: 2024-03-31.

\bibitem[Guo et~al.(2017)Guo, Pleiss, Sun, and Weinberger]{guo2017calibration}
Chuan Guo, Geoff Pleiss, Yu Sun, and Kilian~Q Weinberger.
\newblock On calibration of modern neural networks.
\newblock In \emph{International conference on machine learning}, pages 1321--1330. PMLR, 2017.

\bibitem[Hanley and McNeil(1982)]{hanley1982meaning}
James~A Hanley and Barbara~J McNeil.
\newblock The meaning and use of the area under a receiver operating characteristic (roc) curve.
\newblock \emph{Radiology}, 143\penalty0 (1):\penalty0 29--36, 1982.

\bibitem[He et~al.(2016)He, Zhang, Ren, and Sun]{he2016deep}
Kaiming He, Xiangyu Zhang, Shaoqing Ren, and Jian Sun.
\newblock Deep residual learning for image recognition.
\newblock In \emph{Proceedings of the IEEE conference on computer vision and pattern recognition}, pages 770--778, 2016.

\bibitem[Hinton et~al.(2012)Hinton, Srivastava, Krizhevsky, Sutskever, and Salakhutdinov]{hinton2012improving}
Geoffrey~E Hinton, Nitish Srivastava, Alex Krizhevsky, Ilya Sutskever, and Ruslan~R Salakhutdinov.
\newblock Improving neural networks by preventing co-adaptation of feature detectors.
\newblock \emph{arXiv preprint arXiv:1207.0580}, 2012.

\bibitem[Horien et~al.(2021)Horien, Noble, Greene, Lee, Barron, Gao, O’Connor, Salehi, Dadashkarimi, Shen, et~al.]{horien}
Corey Horien, Stephanie Noble, Abigail~S Greene, Kangjoo Lee, Daniel~S Barron, Siyuan Gao, David O’Connor, Mehraveh Salehi, Javid Dadashkarimi, Xilin Shen, et~al.
\newblock A hitchhiker’s guide to working with large, open-source neuroimaging datasets.
\newblock \emph{Nature human behaviour}, 5\penalty0 (2):\penalty0 185--193, 2021.

\bibitem[Hossain et~al.(2022)Hossain, Iqbal, Islam, Akhtar, and Sarker]{hossain2022transfer}
Md~Belal Hossain, SM~Hasan~Sazzad Iqbal, Md~Monirul Islam, Md~Nasim Akhtar, and Iqbal~H Sarker.
\newblock Transfer learning with fine-tuned deep cnn resnet50 model for classifying covid-19 from chest x-ray images.
\newblock \emph{Informatics in Medicine Unlocked}, 30:\penalty0 100916, 2022.

\bibitem[Huang et~al.(2020)Huang, Kothari, Banerjee, Chute, Ball, Borus, Huang, Patel, Rajpurkar, Irvin, et~al.]{huang2020penet}
Shih-Cheng Huang, Tanay Kothari, Imon Banerjee, Chris Chute, Robyn~L Ball, Norah Borus, Andrew Huang, Bhavik~N Patel, Pranav Rajpurkar, Jeremy Irvin, et~al.
\newblock Penet—a scalable deep-learning model for automated diagnosis of pulmonary embolism using volumetric ct imaging.
\newblock \emph{NPJ digital medicine}, 3\penalty0 (1):\penalty0 1--9, 2020.

\bibitem[Jaeger et~al.(2014)Jaeger, Candemir, Antani, W{\'a}ng, Lu, and Thoma]{jaeger2014two}
Stefan Jaeger, Sema Candemir, Sameer Antani, Y{\`\i}-Xi{\'a}ng~J W{\'a}ng, Pu-Xuan Lu, and George Thoma.
\newblock Two public chest x-ray datasets for computer-aided screening of pulmonary diseases.
\newblock \emph{Quantitative imaging in medicine and surgery}, 4\penalty0 (6):\penalty0 475, 2014.

\bibitem[Kingma and Ba(2014)]{kingma2014adam}
Diederik~P Kingma and Jimmy Ba.
\newblock Adam: A method for stochastic optimization.
\newblock \emph{arXiv preprint arXiv:1412.6980}, 2014.

\bibitem[Mello-Thoms and Mello(2023)]{mellothoms}
Claudia Mello-Thoms and Carlos~AB Mello.
\newblock Clinical applications of artificial intelligence in radiology.
\newblock \emph{The British Journal of Radiology}, 96\penalty0 (1150):\penalty0 20221031, 2023.

\bibitem[Nguyen et~al.(2019)Nguyen, Nguyen, Dao, Unnikrishnan, Dhingra, Ravichandran, Satpathy, Raja, and Chua]{nguyen2019deep}
Quang~H Nguyen, Binh~P Nguyen, Son~D Dao, Balagopal Unnikrishnan, Rajan Dhingra, Savitha~Rani Ravichandran, Sravani Satpathy, Palaparthi~Nirmal Raja, and Matthew~CH Chua.
\newblock Deep learning models for tuberculosis detection from chest x-ray images.
\newblock In \emph{2019 26th international conference on telecommunications (ICT)}, pages 381--385. IEEE, 2019.

\bibitem[Park et~al.(2019)Park, Chute, Rajpurkar, Lou, Ball, Shpanskaya, Jabarkheel, Kim, McKenna, Tseng, et~al.]{park2019deep}
Allison Park, Chris Chute, Pranav Rajpurkar, Joe Lou, Robyn~L Ball, Katie Shpanskaya, Rashad Jabarkheel, Lily~H Kim, Emily McKenna, Joe Tseng, et~al.
\newblock Deep learning--assisted diagnosis of cerebral aneurysms using the headxnet model.
\newblock \emph{JAMA network open}, 2\penalty0 (6):\penalty0 e195600--e195600, 2019.

\bibitem[Pianykh et~al.(2020)Pianykh, Langs, Dewey, Enzmann, Herold, Schoenberg, and Brink]{pianykh}
Oleg~S Pianykh, Georg Langs, Marc Dewey, Dieter~R Enzmann, Christian~J Herold, Stefan~O Schoenberg, and James~A Brink.
\newblock Continuous learning ai in radiology: implementation principles and early applications.
\newblock \emph{Radiology}, 297\penalty0 (1):\penalty0 6--14, 2020.

\bibitem[Rajpurkar et~al.(2017)Rajpurkar, Irvin, Zhu, Yang, Mehta, Duan, Ding, Bagul, Langlotz, Shpanskaya, et~al.]{rajpurkar2017chexnet}
Pranav Rajpurkar, Jeremy Irvin, Kaylie Zhu, Brandon Yang, Hershel Mehta, Tony Duan, Daisy Ding, Aarti Bagul, Curtis Langlotz, Katie Shpanskaya, et~al.
\newblock Chexnet: Radiologist-level pneumonia detection on chest x-rays with deep learning.
\newblock \emph{arXiv preprint arXiv:1711.05225}, 2017.

\bibitem[Rajpurkar et~al.(2020)Rajpurkar, Park, Irvin, Chute, Bereket, Mastrodicasa, Langlotz, Lungren, Ng, and Patel]{rajpurkar2020appendixnet}
Pranav Rajpurkar, Allison Park, Jeremy Irvin, Chris Chute, Michael Bereket, Domenico Mastrodicasa, Curtis~P Langlotz, Matthew~P Lungren, Andrew~Y Ng, and Bhavik~N Patel.
\newblock Appendixnet: Deep learning for diagnosis of appendicitis from a small dataset of ct exams using video pretraining.
\newblock \emph{Scientific reports}, 10\penalty0 (1):\penalty0 1--7, 2020.

\bibitem[Rao et~al.(2021)Rao, Park, Woo, Lee, and Aalami]{rao2021studying}
Adrit Rao, Jongchan Park, Sanghyun Woo, Joon-Young Lee, and Oliver Aalami.
\newblock Studying the effects of self-attention for medical image analysis.
\newblock In \emph{Proceedings of the IEEE/CVF International Conference on Computer Vision}, pages 3416--3425, 2021.

\bibitem[Rao et~al.(2023)Rao, Lee, and Aalami]{rao2023studying}
Adrit Rao, Joon-Young Lee, and Oliver Aalami.
\newblock Studying the impact of augmentations on medical confidence calibration.
\newblock In \emph{Proceedings of the IEEE/CVF International Conference on Computer Vision}, pages 2462--2472, 2023.

\bibitem[Saporta et~al.(2022)Saporta, Gui, Agrawal, Pareek, Truong, Nguyen, Ngo, Seekins, Blankenberg, Ng, et~al.]{saporta2022benchmarking}
Adriel Saporta, Xiaotong Gui, Ashwin Agrawal, Anuj Pareek, Steven~QH Truong, Chanh~DT Nguyen, Van-Doan Ngo, Jayne Seekins, Francis~G Blankenberg, Andrew~Y Ng, et~al.
\newblock Benchmarking saliency methods for chest x-ray interpretation.
\newblock \emph{Nature Machine Intelligence}, 4\penalty0 (10):\penalty0 867--878, 2022.

\bibitem[Selvaraju et~al.(2017)Selvaraju, Cogswell, Das, Vedantam, Parikh, and Batra]{selvaraju2017grad}
Ramprasaath~R Selvaraju, Michael Cogswell, Abhishek Das, Ramakrishna Vedantam, Devi Parikh, and Dhruv Batra.
\newblock Grad-cam: Visual explanations from deep networks via gradient-based localization.
\newblock In \emph{Proceedings of the IEEE international conference on computer vision}, pages 618--626, 2017.

\bibitem[Showkat and Qureshi(2022)]{showkat2022efficacy}
Sadia Showkat and Shaima Qureshi.
\newblock Efficacy of transfer learning-based resnet models in chest x-ray image classification for detecting covid-19 pneumonia.
\newblock \emph{Chemometrics and Intelligent Laboratory Systems}, 224:\penalty0 104534, 2022.

\bibitem[Smailagic et~al.(2018)Smailagic, Costa, Noh, Walawalkar, Khandelwal, Galdran, Mirshekari, Fagert, Xu, Zhang, et~al.]{smailagic2018medal}
Asim Smailagic, Pedro Costa, Hae~Young Noh, Devesh Walawalkar, Kartik Khandelwal, Adrian Galdran, Mostafa Mirshekari, Jonathon Fagert, Susu Xu, Pei Zhang, et~al.
\newblock Medal: Accurate and robust deep active learning for medical image analysis.
\newblock In \emph{2018 17th IEEE international conference on machine learning and applications (ICMLA)}, pages 481--488. IEEE, 2018.

\bibitem[Van~Cleeff et~al.(2005)Van~Cleeff, Kivihya-Ndugga, Meme, Odhiambo, and Klatser]{vancleeff}
MRA Van~Cleeff, LE Kivihya-Ndugga, H Meme, JA Odhiambo, and PR Klatser.
\newblock The role and performance of chest x-ray for the diagnosis of tuberculosis: a cost-effectiveness analysis in nairobi, kenya.
\newblock \emph{BMC infectious diseases}, 5:\penalty0 1--9, 2005.

\bibitem[Winkler et~al.(2019)Winkler, Fink, Toberer, Enk, Deinlein, Hofmann-Wellenhof, Thomas, Lallas, Blum, Stolz, et~al.]{winkler2019association}
Julia~K Winkler, Christine Fink, Ferdinand Toberer, Alexander Enk, Teresa Deinlein, Rainer Hofmann-Wellenhof, Luc Thomas, Aimilios Lallas, Andreas Blum, Wilhelm Stolz, et~al.
\newblock Association between surgical skin markings in dermoscopic images and diagnostic performance of a deep learning convolutional neural network for melanoma recognition.
\newblock \emph{JAMA dermatology}, 155\penalty0 (10):\penalty0 1135--1141, 2019.

\bibitem[Yang et~al.(2017)Yang, Zhang, Chen, Zhang, and Chen]{yang2017suggestive}
Lin Yang, Yizhe Zhang, Jianxu Chen, Siyuan Zhang, and Danny~Z Chen.
\newblock Suggestive annotation: A deep active learning framework for biomedical image segmentation.
\newblock In \emph{Medical Image Computing and Computer Assisted Intervention- MICCAI 2017: 20th International Conference, Quebec City, QC, Canada, September 11-13, 2017, Proceedings, Part III 20}, pages 399--407. Springer, 2017.

\bibitem[Yun et~al.(2019)Yun, Han, Oh, Chun, Choe, and Yoo]{yun2019cutmix}
Sangdoo Yun, Dongyoon Han, Seong~Joon Oh, Sanghyuk Chun, Junsuk Choe, and Youngjoon Yoo.
\newblock Cutmix: Regularization strategy to train strong classifiers with localizable features.
\newblock In \emph{Proceedings of the IEEE/CVF international conference on computer vision}, pages 6023--6032, 2019.

\bibitem[Zech et~al.(2018)Zech, Badgeley, Liu, Costa, Titano, and Oermann]{zech2018variable}
John~R Zech, Marcus~A Badgeley, Manway Liu, Anthony~B Costa, Joseph~J Titano, and Eric~Karl Oermann.
\newblock Variable generalization performance of a deep learning model to detect pneumonia in chest radiographs: a cross-sectional study.
\newblock \emph{PLoS medicine}, 15\penalty0 (11):\penalty0 e1002683, 2018.

\bibitem[Zhang et~al.(2017)Zhang, Cisse, Dauphin, and Lopez-Paz]{zhang2017mixup}
Hongyi Zhang, Moustapha Cisse, Yann~N Dauphin, and David Lopez-Paz.
\newblock mixup: Beyond empirical risk minimization.
\newblock \emph{arXiv preprint arXiv:1710.09412}, 2017.

\end{thebibliography}
}


\end{document}